\documentstyle[preprint,aps,epsfig]{revtex}

\def\lsim{\raise0.3ex\hbox{$\;<$\kern-0.75em\raise-1.1ex
\hbox{$\sim\;$}}}
\def\gsim{\raise0.3ex\hbox{$\;>$\kern-0.75em\raise-1.1ex
\hbox{$\sim\;$}}}
\begin{document}

\draft

\preprint{}

\title{Dark Matter Neutrinos Must Come with Degenerate Masses} 

\author{Hisakazu Minakata\footnote{Email: minakata@phys.metro-u.ac.jp}
 and Osamu Yasuda\footnote{Email: yasuda@phys.metro-u.ac.jp}}

\address{\qquad \\Department of Physics,
Tokyo Metropolitan University \\
Minami-Osawa, Hachioji, Tokyo 192-0397, Japan}

\date{February, 1997}
\preprint{
\parbox{5cm}{
TMUP-HEL-9713\\
%\today\\
hep-ph/9712291\\
}}
\maketitle

\begin{abstract}
It has been known that there are two schemes in the framework of three
flavor neutrinos to accommodate the global features of the hot dark
matter neutrinos, the solar neutrino deficit and the atmospheric
neutrino anomaly in a manner consistent with terrestrial neutrino
experiments, i.e., hierarchical mass neutrinos and almost degenerate
neutrinos.  We demonstrate that the recent result by the CHOOZ
experiment excludes the scheme of hierarchical neutrinos. We also point
out in the scheme of almost degenerate neutrinos that if neutrinos are
Majorana particles then the double $\beta$ decay experiments must see
positive signals on their way to reach a limit more stringent than the
present one by a factor of 5.

\end{abstract}

\newpage

\section{Introduction}
Massive neutrinos are the only known particle candidates for the 
dark matter in the universe \cite{KT}. While it is unlikely that 
they are responsible for the dark matter as a whole in the cosmos, 
they serve as the best candidates for the hot component in the mixed 
(cold + hot) dark matter model \cite {chdm}. 
The model is one of the rival models that can account for the observed 
power spectrum of density fluctuations over wide length scales, from 
galaxy-galaxy to much longer scales probed by the COBE observation 
\cite {WS}. 

The remaining unresolved issue in the model is how many flavors 
of neutrinos are acting as dark matter, e.g., only the tau 
neutrino or all three of them. The parameter $\Omega_{\nu}$ 
\cite{KT}, the fraction of the density of neutrinos in the 
universe, is sensitive to the sum of neutrino masses but not 
to the neutrinos mass spectrum. 
The problem of which mass pattern of neutrino is favored has been 
discussed based on the goodness of fit to various cosmological 
observables \cite {chdm}. However, the definitive conclusion 
does not appear to be reached yet. 

Here we make a different approach to the same problem based
on a particle physics point of view. We demand that our scheme of the
hot dark matter neutrinos satisfy the following
requirements in the framework of three-flavor neutrinos : \\
\noindent
(1) Neutrinos exist in the standard three-flavor mixing framework 
without sterile species. \\
\noindent
(2) The three-flavor mixing framework must accommodate the deficit 
of the solar neutrinos \cite{solar} and the anomaly in the ratio 
$R = \displaystyle\frac{(\nu_{\mu}/\nu_e)_{\rm observed}}
{(\nu_{\mu}/\nu_e)_{\rm expected}}$ observed in the atmospheric 
neutrino experiments \cite{atmospheric}.\\
\noindent
(3) The three-flavor mixing framework must be consistent with the
constraints imposed by all the reactor and the accelerator
experiments.\footnote{ In this paper, we take a conservative attitude
to the experimental results obtained by the LSND group \cite {LSND}
who claims to observe the appearance signals consistent with neutrino
oscillations. We feel that we should wait for confirmation by other
experiments, KARMEN and Fermilab, before concluding it as an evidence
for neutrino oscillations. The similar attitude is shared by the
author of Ref. \cite {Hill} }

Several remarks are in order on the requirements (1)-(3). \\
\noindent
(i) The dark matter neutrino hypothesis requires neutrino masses but not 
flavor mixings. However, it is a legitimate assumption in particle physics 
that neutrino's gauge and mass eigenstates differ if they are massive. 
It is a natural assumption because we know that the flavor mixing occurs 
in the quark sector \cite{PDG}. \\
\noindent
(ii) It appears that the existence of the solar neutrino deficit and
the atmospheric neutrino anomaly are robust, as confirmed by the
recent high-statistics measurements by Superkamiokande \cite{totsuka}. 
The most likely explanation of these anomalies is due to the neutrino
masses and the flavor mixing. \\
\noindent
(iii) Among other terrestrial experiments the recent $\bar{\nu_e}$ 
disappearance experiment done by the CHOOZ group \cite{CHOOZ} puts
stringent constraints on the three-flavor framework we take 
in this paper. 

In the framework of the three flavor neutrinos two schemes of the hot
dark matter neutrinos have been known which satisfy the requirements
(1) -- (3).  One is the scheme of almost degenerate neutrinos (ADN)
\cite {ADN}, in which the mass squared differences $\Delta m^2$ of
neutrinos are much smaller than several eV$^2$.  The other one is that
of hierarchical mass neutrinos (HMN) \cite {mina1,AP,GKR}, in which at
least one of $\Delta m^2$ reflects the dark matter mass scale, $\sim$
a few eV.  In this paper we show that the recent result by the CHOOZ
experiment excludes the HMN scheme.  We also show that the CHOOZ
result puts a strong constraint on the ADN scheme and that the ADN
scenario predicts a positive signal in neutrinoless double $\beta$
decay experiments in the near future.

In section 2 we summarize the constraints from the accelerator and
reactor experiments.  In section 3 we perform a qualitative analysis
in which we only rely on the global features of the anomalies. 
In section 4 we analyze
in detail the goodness of fit of the HMN scheme to the solar and the
atmospheric neutrino data, taking into account more sophisticated
issues, such as the zenith-angle distribution of the atmospheric
neutrino data.  In section 5 we discuss the constraints on the hot
dark matter scenario with the ADN scheme by the CHOOZ result and the
neutrinoless double $\beta$ decay experiments.  In section 6 we
summarize our conclusions and give remarks on possible signals in the
near future experiments.

\section{Constraints from Accelerator and the Reactor Experiments}
The mass-squared differences $\Delta m^2$ greater than 1 eV$^2$ are
subject to the constraints by the short-baseline accelerator and the
reactor experiments. Therefore, the mixing parameters of the HMN
scenario are tightly constrained by them. This point has recently been
revealed explicitly by the authors of Refs.\cite{mina1,mina2,BBGK,FLS}
in a generic three-flavor mixing framework, whereas it may have been
implicitly noticed before. In particular, the resulting constraint is
quantitatively worked out by Fogli, Lisi and Scioscia \cite{FLS}.

To make the discussion clearer we take a particular representation of
the neutrino mixing matrix $U$, which connects the gauge and the mass
eigenstates as $\nu_{\alpha} = U_{\alpha i}\nu_i \quad
(\alpha = e,\mu, \tau. i=1,2,3)$,
\begin{eqnarray}
U=\left[
\matrix {c_{12}c_{13} & s_{12}c_{13} &  s_{13}e^{-i\delta}\nonumber\\
-s_{12}c_{23}-c_{12}s_{23}s_{13}e^{i\delta} &
c_{12}c_{23}-s_{12}s_{23}s_{13}e^{i\delta} & s_{23}c_{13}\nonumber\\
s_{12}s_{23}-c_{12}c_{23}s_{13}e^{i\delta} &
-c_{12}s_{23}-s_{12}c_{23}s_{13}e^{i\delta} & c_{23}c_{13}\nonumber\\}
\right].
\label{CKM}
\end{eqnarray}

The constraint by the short-baseline experiments is powerful enough 
to allow only the three regions on plane spanned by mixing angles 
$\theta_{13}$ and $\theta_{23}$. 

\noindent
(a) small-s$_{13}$ and small-s$_{23}$\\
\noindent
(b) large-s$_{13}$ and arbitrary s$_{23}$\\
\noindent
(c) small-s$_{13}$ and large-s$_{23}$,\\
Each region can be characterized by approximate decoupling of a neutrino 
state, $\nu_{\tau}, \nu_e$, and $\nu_{\mu}$ for (a), (b) and (c) 
respectively. If we (naturally) allow the remaining angle $\theta_{12}$ 
to be large we have large-angle mixing between $\nu_{\mu} \leftrightarrow 
\nu_e, \nu_{\mu} \leftrightarrow \nu_{\tau}$, and 
$\nu_e \leftrightarrow \nu_{\tau}$ 
in regions (a), (b) and (c), respectively. 
Since we attempt to implement the atmospheric neutrino anomaly into our 
framework we cannot live in the region (c). 
It also implies that in the regions (a) and (b) the atmospheric neutrino 
anomaly is due to almost pure $\nu_{\mu} \leftrightarrow \nu_e$ and 
$\nu_{\mu} \leftrightarrow \nu_{\tau}$ oscillations, respectively. 

In summary, HMN scenario is strongly constrained by the short-baseline
experiments but ADN is free from such constraints.\footnote{ A mild
restriction on $\theta_{13}$ in ADN scenario arises from the Bugey
\cite{Bugey} and the Krasnoyarsk \cite{Krasnoyarsk} experiments if the
scale of $\Delta m^2$ relevant to the atmospheric neutrino anomaly
falls in region $\Delta m^2 \gsim 10^{-2}$ eV$^2$.}

The recent CHOOZ experiment has drastically altered the above
situation. The experiment measured the $\bar{\nu_e}$ beam attenuation
from the reactors by the Gd-loaded liquid scintillator detector
located at about 1 km from the reactors. Due to the long-baseline of
the experiment they are able to probe the wide region of $\Delta m^2
\gsim 10^{-3}$eV$^2$, which essentially cover most of the
region relevant to the atmospheric neutrino anomaly. The negative
result in the CHOOZ experiment put stringent constraints on the mixing
parameters as we will discuss below.
 
Let us discuss the constraints on the mixing parameters implied by the
CHOOZ result.  We start by describing the simplification that occurs
to the neutrino oscillation probabilities under the mass hierarchy,
which will play an important role in the following discussions. For
definiteness of the notation we deal with the mass hierarchy 
depicted in Fig. 1. Namely, the
larger mass squared difference $\Delta m^2_{31} \simeq \Delta m^2_{32}$,
which we denote as $\Delta M^2$ in this paper, is much greater than
the smaller one, $\Delta m^2 \equiv \Delta m^2_{21}$. If the neutrino
masses obey the hierarchy $\Delta M^2 \gg \Delta m^2$, the
disappearance probability can be approximately written as
\begin{equation}
1-P(\bar{\nu_{\alpha}} \rightarrow \bar{\nu_{\alpha}})
= 4|U_{\alpha 3}|^2 (1-|U_{\alpha 3}|^2)\sin^2
\left(\frac{\Delta M^2 L}{4E}\right)
+ 4|U_{\alpha 1}|^2 |U_{\alpha 2}|^2\sin^2
\left(\frac{\Delta m^2 L}{4E}\right).
\label{disappear}
\end{equation}
The correction terms to (\ref{disappear}) are negligible provided 
that $\Delta M^2 L/4E$ is so large that oscillations due to 
$\sin (\Delta M^2 L/4E)$ is averaged to zero, or 
$\sin (\Delta m^2 L/4E)$ is sufficiently small. 
A convenient formula to estimate these quantities is: 
\[
\displaystyle\frac{\Delta m^2 L}{4E}  =
1.27 \left(\displaystyle\frac{\Delta m^2}{10^{-3}\mbox{eV}^2}\right)
\left(\frac{L}{1\mbox{km}}\right)\left(\frac{E}
{1\mbox{MeV}}\right)^{-1}.
\]

We first discuss the constraints imposed on the ADN scenario. In this
scenario role of the dark matter is played by all the three neutrinos
with almost equal masses of the order of a few eV, with the mass
difference roughly of the order of $\Delta M^2 = \Delta m^2_{\rm atm}
\simeq 10^{-3} - 10^{-2}$ eV$^2$.
The smaller $\Delta m^2$ depends upon whether we take the MSW solution
of the solar neutrino problem \cite {MSW} for which $\Delta m^2 =
\Delta m^2_{\rm solar} \simeq 10^{-6} - 10^{-5}$ eV$^2$,
or the just-so vacuum oscillation solution \cite {just-so}, $\Delta
m^2 \simeq 10^{-10}$ eV$^2$. For notational convenience, we use in the
rest of the paper the symbols $\Delta m^2_{\rm atm}$ and $\Delta
m^2_{\rm solar}$, which stands for the above mass scales relevant to
the atmospheric neutrino anomaly and the solar neutrino deficit,
respectively.

Since the neutrino oscillation phenomenon is sensitive only to 
the squared-mass difference, not to the absolute mass values, 
the following discussion of the constraints on the ADN case 
also applies to the case of no dark matter mass neutrinos, e.g., 
$m_1 \ll m_2 \sim 10^{-3}- 10^{-2}$eV, and $m_3 \sim 0.03-0.1$eV, 
to which we will refer as ELN (extremely light neutrinos).

In ADN type $\Delta m^2$-hierarchy the disappearing rate in 
$\bar{\nu_e} \rightarrow \bar{\nu_e}$ experiment can be written as
\begin{equation}
1-P(\bar{\nu_e} \rightarrow \bar{\nu_e}) 
= 4c_{13}^2s_{13}^2\sin^2\left(\displaystyle
\frac{\Delta M^2 L}{4E}\right), \nonumber
\end{equation}
because the second term in (\ref{disappear}) is negligible for $\Delta
m^2 = \Delta m^2_{\rm solar}$. It is easy to observe that the bound
obtained by the CHOOZ group can be easily translated into the one for
$s_{13}$.  We have calculated the allowed region for the
$\bar{\nu_e}\leftrightarrow\bar{\nu_e}$ disappearance experiments at
90 \% confidence level by combining the CHOOZ, the older Bugey and the
Krasnoyarsk results. The result is presented in Fig.2.

\section{Qualitative Analysis of the HMN scheme}

We perform a qualitative analysis of the HMN scenario in which 
we rely on global features of the anomalies of the solar and the 
atmospheric neutrino data. Namely, we require for solar neutrinos 
the suppression of the flux by at least factor of 2. 
It is a milder requirement than requiring the consistency with 
the varying suppression rate observed in the different experiments 
which seem to indicate that the depletion rate of the solar 
neutrino flux is strongly energy-dependent. Similarly, we only 
require the consistency with the gross deviation of the ratio $R$ 
in the atmospheric neutrino data from unity.  

In the HMN scenario $\Delta M^2 = \Delta m^2_{\rm DM} \sim$ several
eV$^2$, and $\Delta m^2 = \Delta m^2_{\rm atm}$ or $\Delta m^2 = \Delta
m^2_{\rm solar}$ as required by the condition (2). If we exploit the
second option for $\Delta m^2$ in order to incorporate naturally the
solar neutrino deficit, there is no way of accommodating the
atmospheric neutrino anomaly.\footnote{ There is a marginal solution
that can take into account of the atmospheric neutrino anomaly with
the choice of $\Delta m^2 = \Delta m^2_{\rm solar}$, as proposed by
Cardall and Fuller \cite{CF} and recently confirmed by more elaborate
analysis by Fogli et al.\cite{FLMS}. However, it is questionable
\cite{MY1} if the solution survives after the zenith angle
distribution is taken into account. Furthermore, the solution survives
only with a small value of $\Delta M^2 \simeq 0.45$ eV$^2$, a too
small value for the model of hierarchical mass neutrino as hot dark
matter, while it is consistent with the LSND data from which their
original motivation comes.  } This point is in fact well known and is
recently summarized in Ref.\cite{mina3}. It will be discussed more
quantitatively below. For this reason we employ the first option,
$\Delta m^2 = \Delta m^2_{\rm atm} \simeq 10^{-3} - 10^{-2}$ eV$^2$. This
mass hierarchy has recently been discussed in Refs.
\cite {mina1,AP,GKR}.

Under the mass spectrum we just specified the arguments
of the first and the second sine functions in (\ref{disappear}) are of
order unity for the short- and the long-baseline experiments,
respectively, if the neutrino beam energy is of order $\sim$1 GeV. 
Then, the short- and the long-baseline neutrino experiments can give
rise to independent different constraints on mixing parameters. Let
us discuss it more explicitly by restricting to $\bar{\nu_e}$
disappearance experiment. With use of the definition of the mixing
angles in (\ref{CKM}) the $\bar{\nu_e}$ disappearance probability at
long-baseline can be written as
\begin{equation}
1-P(\bar{\nu_e}\rightarrow \bar{\nu_e}) =2s^2_{13}c^2_{13} + 
4c^2_{12}s^2_{12}c^4_{13}\sin^2 \left(\displaystyle\frac{\Delta m^2 L}{4E}
\right).
\label{eqn:pee}
\end{equation}
To indicate the point let us take $\Delta M^2=5$ eV$^2$. 
Then, the parameter $s^2_{13}c^2_{13}$ is constrained to be 
$s^2_{13}c^2_{13} \leq 2 \times10^{-2}$ and thus the first term 
in (\ref{disappear}) is smaller than $4\times 10^{-2}$. 
Therefore, the long-baseline $\bar{\nu_e}$ disappearance 
experiment primarily constrains the angle $\theta_{12}$. 

Let us discuss the parameter region (a) and (b) separately. 

\noindent
The region (a): small $s_{13}$ and small $s_{23}$

\noindent
In this region $c_{13}$ is approximately unity. Then, the neutrino 
disappearance probability is effectively of the type of two-flavor mixing. 
One can translate the bound on $\sin^2 2\theta$ obtained by the CHOOZ group 
to that on $\sin^2 2\theta_{12}$. 
The most conservative bound is obtained for $s^2_{12}$ by ignoring 
the first term of (\ref{disappear}) and it coincides with the one 
obtained for $s^2_{13}$ in the preceding analysis of the ADN scenario. 

\noindent
The region (b): large $s_{13}$ and arbitrary $s_{23}$

\noindent 
In this region $c_{13}$ is small, and the Bugey experiment \cite{Bugey}
implies $c^2_{13} \leq 2\times10^{-2}$.  If we take the smaller
mass squared difference $\Delta m^2 = 2\times 10^{-3}$eV$^2$ then
the CHOOZ experiment yields a little stronger
constraint on $c_{13}$ in this region (See section 4).

We now address the messages that are most important in this paper; 
In the case of HMN the gross deficit of the solar neutrino flux cannot 
be achieved in any of the parameter regions (a) and (b). 
In the region (a) because of the bound on $s^2_{12}$ the solar neutrino 
deficit is at most $\sim$20 \% if $\Delta m^2_{\rm atm} \geq 2\times 
10^{-3}$ eV$^2$. We will address the case of smaller 
$\Delta m^2_{\rm atm}$ in our quantitative analysis later. 

In the region (b) the solar neutrino deficit does not occur as was
pointed out in Ref. \cite{mina2,BBGK}. 
One can easily understand this by noticing the approximate expression 
of the three-flavor oscillation probability,  
\begin{eqnarray}
P(\nu_e \rightarrow \nu_e) &=& 1 - 2(c^4_{13}s^2_{12}c^2_{12}
+s^2_{13}c^2_{13})\nonumber\\
&\simeq&1,\nonumber
\end{eqnarray}
which is obtained from (\ref{eqn:pee}) after taking average over rapid
oscillations, since the smallest mass squared difference $\Delta m^2$
is much larger than $\Delta m^2_{\rm solar}$.  It implies that the
large $s_{13}$ region (b) cannot accommodate the gross deficit of
solar neutrino flux.

Even worse one cannot have enough atmospheric neutrino anomaly in 
the region (a). To achieve a qualitative understanding of this fact 
prior to the quantitative analysis pursued later we ignore the matter 
effect and write down the oscillation 
probability at long-baseline as 
\begin{eqnarray}
P(\nu_{\mu} \rightarrow \nu_e) && \nonumber\\
&=& 2c^2_{13}s^2_{13}s^2_{23}
+ \left[4c^2_{12}s^2_{12}c^2_{13}(c^2_{23}-s^2_{23}s^2_{13}) 
+ 4(c^2_{12}-s^2_{12}) J \cot\delta \right]
\sin^2 \left(\displaystyle\frac{\Delta m^2L}{4E}\right)\nonumber\\
&& + 2J\sin\left(\frac{\Delta m^2 L}{2E}\right),\nonumber
\end{eqnarray}
where $J=c_{12}s_{12}c^2_{13}s_{13}c_{23}s_{23}\sin\delta$ denotes 
the leptonic Jarlskog factor. At $\Delta M^2=5$eV$^2$ the first 
term is small, $\leq 1.6\times 10^{-4}$. 
The third term is also small, $\leq 5\times 10^{-3}$. 
By imposing the bound obtained by CHOOZ group \cite {CHOOZ} on 
the coefficient of the second term one can show that it is 
smaller than 0.21. 
Because of the dominance of $\nu_{\mu} \rightarrow \nu_e$ oscillation 
we obtain, as a rough estimation of the ratio $R$, 
$R= (1-0.1)/(1+0.1) \simeq 0.8$, which may not be sufficient to explain 
the data of atmospheric neutrino observation. 

This concludes our qualitative discussions to demonstrate that 
the HMN dark matter cannot be reconciled with the requirements 
(2) and (3). 
We therefore conclude that the dark matter neutrinos must come with 
almost degenerate mass spectrum. 

\section{Quantitative analysis of the HMN scheme}
We quantify our discussion by contrasting the various hypothesis
against the experimental data.  For the regions (a) -- (c), we
optimize the combined $\chi^2$:
\begin{eqnarray}
\chi^2_{\rm tot}\equiv\chi^2_{\rm solar}+\chi^2_{\rm atm}\nonumber,
\end{eqnarray}
where $\chi^2_{\rm solar}$ and $\chi^2_{\rm atm}$ are the $\chi^2$ for
the solar and the atmospheric neutrino data, respectively.  To make
our discussions concrete, let us first take the smaller mass squared
difference $\Delta m^2 = 2\times 10^{-3}$eV$^2$.  We will discuss the
case for $\Delta m^2 \ne 2\times 10^{-3}$eV$^2$ later.

To compute $\chi^2_{\rm solar}$ we adopt the theoretical predictions
by \cite{BP95} and the solar neutrino data quoted in the table in
\cite{HL} with the 6 day-night bins of the Kamiokande data
\cite{Kam} (\# of the data = 6 (Kamiokande) + 1 (Superkamiokande)
+ 1 (Cl) + 1 (Ga, combined) = 9) and we minimize the $\chi^2_{\rm
solar}$ by varying the suppression probability
$P(\nu_e\rightarrow\nu_e)$ and the normalization factor $f_B$ of the
${}^8$B neutrino flux as in Ref. \cite{AP}\footnote{The most recent
solar neutrino data of Superkamiokande and GALLEX are slightly
different from those quoted in
\cite{HL}, but it turns out that the conclusions below do not change.
Also our conclusions below do not change even if we adopt other
theoretical predictions \cite{BP92,TL,DS} for solar neutrinos.}.  

The theoretical predictions to the atmospheric neutrino data are
computed by using a code developed by one of the authors
\cite{yasuda}.  We perform separate analyses of the Kamiokande
\cite{fukuda} and the Superkamiokande data \cite{totsuka} of
atmospheric neutrino observation. We use the Kamiokande atmospheric
neutrino data (\# of the data = 10 (multi-GeV, zenith angle) + 5
(sub-GeV, zenith angle) + 20 (sub-GeV, energy spectrum) = 35) and the
Superkamiokande data (\# of the data = 10 (multi-GeV, zenith angle) +
10 (sub-GeV, zenith angle) + 25 (sub-GeV, energy spectrum) = 45) in
the following.

In the region (a) we have a poor fit to both the solar and atmospheric
neutrino data.  The most stringent constraint on $s^2_{13}$ comes from
the CHOOZ experiment \cite{CHOOZ}.  Namely, using the disappearance
probability (\ref{eqn:pee}) with $\Delta m^2=2\times10^{-3}$eV$^2$,
the allowed region at 90 \% confidence level can be obtained as
$s^2_{13} \le 3\times10^{-2}$ and $\sin^2 2\theta_{12} \le 0.4$.  The
most stringent bound on $s^2_{23}$ comes from the E531 experiment
($\nu_\mu\rightarrow\nu_\tau$ appearance) \cite{E531}: $s^2_{23}\le
1\times10^{-3}$.  Since we would like to demonstrate that our argument
also holds for all the regions of $\Delta M^2 >$ 5 eV$^2$, we use less
severe constraint from the CDHSW experiment
($\nu_\mu\leftrightarrow\nu_\mu$ disappearance)
\cite{CDHSW} here: $s^2_{23}\le 2\times10^{-2}$.  For $\Delta M^2 >$ 5
eV$^2$ the bound on $s^2_{23}$ from the E531 experiment \cite{E531}
becomes stronger.  Under these constraints we obtain
\begin{eqnarray}
\chi^2_{\rm solar}&\ge&81.3\quad({\rm 9~d.f.})\nonumber\\
\chi^2_{\rm atm}&\ge&\left\{ \begin{array}{rrl}
105.8&\quad&({\rm Kamiokande;~35~d.f.})\nonumber\\
175.8&\quad&({\rm Superkamiokande;~45~d.f.}),
\end{array} \right.\nonumber
\end{eqnarray}
so that we have
\begin{eqnarray}
\chi^2_{\rm tot}\ge\left\{ \begin{array}{rrl}
187.1&\quad&({\rm with~Kamiokande~atm.~\nu;~38~d.f.})\\
257.1&\quad&({\rm with~Superkamiokande~atm.~\nu;~48~d.f.})
\end{array} \right\}~~{\rm for~region~(a)},
\label{eqn:chitot}
\end{eqnarray}
where we have subtracted the number (=6) of degrees of freedom of the
parameters ($\Delta m^2_{21}$, $\Delta m^2_{32}$, $\theta_{12}$,
$\theta_{13}$, $\theta_{23}$, $f_B$) in (\ref{eqn:chitot}).  We note
in passing that the reason that a fit to the Superkamiokande
atmospheric neutrino data is much worse is because our scheme (a)
gives dominant $\nu_\mu\leftrightarrow\nu_e$ oscillations which are
disfavored by the Superkamiokande data.  Thus we conclude that the
region (a) is excluded at ($1-1\times10^{-21}$) CL (or 9.5 standard
deviation) in case of $\chi^2_{\rm solar}+\chi^2_{\rm
atm}$(Kamiokande) and at ($1-2\times10^{-30}$) CL (or 11.5 standard
deviation) in case of $\chi^2_{\rm solar}+\chi^2_{\rm
atm}$(Superkamiokande), respectively.

In the region (b) we have a poor fit to the solar neutrino data.  The
CHOOZ experiment again gives the strongest bound on $\theta_{13}$ and
$c^2_{13} \le 1\times10^{-2}$.  With this condition we find
\begin{eqnarray}
\chi^2_{\rm solar}\ge 127.8\quad({\rm 9~d.f.}).\nonumber
\end{eqnarray}
The fit to the atmospheric neutrino data is moderate:
\begin{eqnarray}
\chi^2_{\rm atm}&\ge&\left\{ \begin{array}{rrl}
47.4&\quad&({\rm Kamiokande;~35~d.f.})\nonumber\\
78.2&\quad&({\rm Superkamiokande;~45~d.f.}).
\end{array} \right.\nonumber
\end{eqnarray}
Therefore we have
\begin{eqnarray}
\chi^2_{\rm tot}\ge\left\{ \begin{array}{rrl}
175.2&\quad&({\rm with~Kamiokande~atm.~\nu;~38~d.f.})\\
206.0&\quad&({\rm with~Superkamiokande~atm.~\nu;~48~d.f.})
\end{array} \right\}~~{\rm for~region~(b)},\nonumber
\end{eqnarray}
and we conclude that the region (b) is excluded at
($1-2\times10^{-19}$) CL (or 9.0 standard deviation) in case of
$\chi^2_{\rm solar}+\chi^2_{\rm atm}$(Kamiokande) and at
($1-2\times10^{-21}$) CL (or 9.5 standard deviation) in case of
$\chi^2_{\rm solar}+\chi^2_{\rm atm}$(Superkamiokande), respectively.

In the region (c) a fit to the solar neutrino data is moderate
and a fit to the atmospheric neutrino data is poor.
The severest constraints on $s^2_{13}$ come from the E776 experiment ($\nu_\mu\rightarrow\nu_e$
appearance)\cite{E776} and we have
$s^2_{13} \le 5\times10^{-4}$.  Although
the E531 experiment ($\nu_\mu \rightarrow
\nu_\tau$ appearance)\cite{E531} gives the strongest
bound on $s^2_{23}$, we use less tight bound from the CDHSW experiment
($\nu_\mu\leftrightarrow\nu_\mu$ disappearance) \cite{CDHSW} which
gives $c^2_{23} \le 2\times10^{-2}$.  Under these conditions we get
\begin{eqnarray}
\chi^2_{\rm solar}&\ge& 29.8\quad({\rm 9~d.f.}),\nonumber\\
\chi^2_{\rm atm}&\ge&\left\{ \begin{array}{lr}
74.2\quad({\rm Kamiokande;~35~d.f.})\nonumber\\
144.4\quad({\rm Superkamiokande;~45~d.f.}).
\end{array} \right.
\end{eqnarray}
So we have
\begin{eqnarray}
\chi^2_{\rm tot}\ge\left\{ \begin{array}{rrl}
104.0&\quad&({\rm with~Kamiokande~atm.~\nu;~38~d.f.})\\
174.2&\quad&({\rm with~Superkamiokande~atm.~\nu;~48~d.f.})
\end{array} \right\}~~{\rm for~region~(c)},\nonumber
\end{eqnarray}
and we conclude that the region (c) is excluded at
($1-5\times10^{-8}$) CL (or 5.5 standard deviation) in case of
$\chi^2_{\rm solar}+\chi^2_{\rm atm}$(Kamiokande) and at
($1-3\times10^{-16}$) CL (or 8.2 standard deviation) in case of
$\chi^2_{\rm solar}+\chi^2_{\rm atm}$(Superkamiokande), respectively.
As we have seen above, irrespective of which
atmospheric neutrino data we use, Kamiokande or Superkamiokande,
the regions (a), (b) and (c) are completely excluded for
$\Delta m^2 = 2\times10^{-3}$eV$^2$.

We have also investigated the case for $\Delta m^2 <
2\times10^{-3}$eV$^2$ and for $\Delta m^2 > 2\times10^{-3}$eV$^2$.
For $\Delta m^2 > 2\times10^{-3}$eV$^2$ the CHOOZ result gives even
severer constraint on $\sin^2 2\theta_{12}$, so a fit to the solar and
atmospheric neutrino data becomes even worse for all the regions (a)
-- (c), and hence (a) -- (c) are excluded for $\Delta m^2 >
2\times10^{-3}$eV$^2$.  For $\Delta m^2 < 2\times10^{-3}$eV$^2$ we
have basically no constraint from the CHOOZ result on $s_{12}$, so a
fit to the solar neutrino data becomes good in the region (a).  On the
other hand, the zenith angle dependence cannot be reproduced for the
atmospheric neutrino data as $\Delta m^2$ becomes smaller, so a fit to
the atmospheric neutrino data becomes worse.  We found that
\begin{eqnarray}
\chi^2_{\rm tot}>\left\{ \begin{array}{rrl}
120&\quad&({\rm with~Kamiokande~atm.~\nu;~38~d.f.})\\
220&\quad&({\rm with~Superkamiokande~atm.~\nu;~48~d.f.})
\end{array} \right\}~~{\rm for~region~(a)},\nonumber
\end{eqnarray}
for $\Delta m^2 < 2\times10^{-3}$eV$^2$, and hence the region (a) is
excluded at ($1-2\times10^{-10}$) CL (or 6.4 standard
deviation) in case of $\chi^2_{\rm solar}+\chi^2_{\rm
atm}$(Kamiokande) and at ($1-7\times10^{-24}$) CL (or 10.1 standard
deviation) in case of $\chi^2_{\rm solar}+\chi^2_{\rm
atm}$(Superkamiokande), respectively.  In the regions (b) and (c) we
do not have any improvement on a fit to the solar neutrino data for
$\Delta m^2 < 2\times10^{-3}$eV$^2$, whereas a fit to the atmospheric
neutrino data becomes poorer.  Therefore we can exclude the
regions (b) and (c) also for $\Delta m^2 < 2\times10^{-3}$eV$^2$.

In the discussions above we have taken the larger mass squared
difference $\Delta M^2 = 5$eV$^2$ as a reference value.  For $\Delta
M^2 >$ 5 eV$^2$ the constraint on $s^2_{23}$ from the E531 experiment
becomes even tighter while the bound on $s^2_{13}$ remain the same as
$\Delta M^2 =$ 5 eV$^2$.  A fit to the solar neutrino and atmospheric
neutrino data is as poor as the case for $\Delta M^2 =$ 5 eV$^2$,
because all the factors $\sin^2(\Delta m^2_{ij}L/4E)$ are averaged to
give 1/2.  So our conclusions still hold for $\Delta M^2 >$ 5 eV$^2$.
For 1 eV$^2\lsim\Delta M^2 \lsim$ 5 eV$^2$, the bound on $s^2_{23}$
becomes weaker, but it turns out that a fit to the solar and
atmospheric neutrino data does not improve, since we are considering
the energy independent solution to the solar neutrino problem and the
zenith angle dependence of the atmospheric neutrino data is not
sensitive to the value of the larger mass squared difference $\Delta
M^2$.  Thus our conclusions do not change as long as the larger mass
squared difference satisfies $\Delta M^2\gsim$ 1 eV$^2$, which covers
the region suggested by the mixed dark matter scenario \cite{chdm}.

To summarize, we have shown quantitatively that our hierarchical schemes
in the regions (a) -- (c) are all excluded.

\section{Analysis of the ADN scenario}
We further discuss the meaning of the constraint imposed on the 
ADN scenario by the CHOOZ experiment. We point out that if
neutrinos are Majorana particles the neutrinoless double $\beta$ 
decay experiments further tighten the parameters of the ADN scenario. 
It has been pointed out \cite{PSMN,MY2} that the CP violating
phases have to be large for the ADN scheme to be consistent with
the neutrinoless double $\beta$ decay experiments.  In particular
we have derived in Ref.\cite{MY2} the inequality that is valid in the 
ADN scenario: 
\[
r \equiv \displaystyle\frac{\langle m_{\nu_e}\rangle}{m} \geq 
\left|c^2_{13}\sqrt{1-\sin^2 \beta \sin^2 2\theta_{12}} -s^2_{13}\right|
\]
where $m\equiv m_1\simeq m_2\simeq m_3$ and $\beta$ is an extra
Majorana CP phase. The most conservative bound is obtained for
$\sin\beta =1$.

If $\Delta m^2_{\rm atm} \geq 2\times10^{-3}$eV$^2$ the 
constraint by the CHOOZ experiment implies that $s^2_{13} \leq 0.05$. 
We further assume the MSW mechanism \cite{MSW} as a solution to 
the solar neutrino problem. 
By looking into Fig.1 of Ref.\cite{MY2} 
one can understand if $r < 0.16$ 
there is no consistent solution with the 90\% confidence level 
allowed region of the solar neutrino deficit derived with the 
MSW mechanism \cite{FLM}. 
This implies the bound for dark matter neutrino mass 
\[
m \leq \displaystyle \frac{\langle m_{\nu e}\rangle}{0.16}.
\]
The experimental bound on $\langle m_{\nu e}\rangle$ obtained by 
the Heidelberg-Moscow group \cite {beta}, 
$\langle m_{\nu e}\rangle \leq 0.45$eV \cite {Klapdor} can be 
translated into the constraint on the degenerate neutrino mass as 
$m \leq 2.8$eV. 
If we include a factor of 2 uncertainty of the nuclear matrix 
elements the bound may be relaxed to $m \leq 5.6$eV. These bounds 
can be expressed into the ones for $\Omega_{\nu}$, 
\[
\Omega_{\nu} = \displaystyle\frac{\sum_i m_i}{91.5 \mbox{eV}} h^{-2} 
\leq 0.1(0.2) h^{-2}
\]
without (with) uncertainty of nuclear matrix element, where h 
indicates the Hubble parameter measured in units of 
100 km/s$\cdot$ Mpc. 

If the Heidelberg-Moscow group continues to see no neutrinoless double
$\beta$ decay event in the present setting the constraint on $\langle
m_{\nu e}\rangle$ will be tighten to $\langle m_{\nu e}\rangle \leq
0.1$ \cite {Klapdor}. If this happened, it implies the bound on
neutrino mass $m \leq$ 0.6 eV (1.2 eV) without (with) the extra
uncertainty of the nuclear matrix elements. These lead to
$\Omega_{\nu} \leq 0.02 (0.04) h^{-2}$ which may be too small to meet
the demand by the mixed dark matter cosmology. This leads to the
important conclusion that if the hot dark matter neutrinos are
Majorana particles and they exist in nature in a manner satisfying the
requirements (1)-(3) mentioned in section 1 then the double $\beta$
decay experiments must see positive signals on the way to reach the
sensitivity down to $\langle m_{\nu e}\rangle =0.1$ eV.

\section{Conclusions}
We have shown in the framework of three flavor neutrino oscillations
that all schemes with hierarchical mass scales in which the larger and
the smaller mass squared differences are of order several eV$^2$ and
10$^{-3}$ -- 10$^{-2}$ eV$^2$, respectively are excluded on the firm 
statistical ground of more than 5 $\sigma$.  We have
also shown that the almost degenerate neutrino scenario can be tested
in neutrinoless double $\beta$ decay experiments in the near future if
neutrinos are of the Majorana type.

Finally several remarks are in order:

\noindent
(1) In a previous communication \cite{MS} we have argued that if the
ratio $R_{lb}$ = (No. of observed electron events)/ (No. of expected
muon events - No. of observed muon events) that can be measured by the
long-baseline neutrino experiments falls into the region $0.02 \leq
R_{lb} \leq 0.87 $ (for $\Delta M^2 \geq 5$ eV$^2$) then the
hierarchical mass dark matter neutrino hypothesis can be rejected. We
have shown in this paper that the new constraint imposed by the CHOOZ
experiment and the requirement of accommodating the solar neutrino
deficit implies that it must be the case.

\noindent
(2) On the other hand, if the CHORUS \cite {CHORUS} or the NOMAD \cite
{NOMAD} experiments observe the appearance signals it turns out that
our conclusion is wrong. It then indicates that at least one of our
assumptions (1) - (3) is incorrect. It would imply a strong indication
for unexpected new features in physics of neutrinos.

\noindent
(3) In general the new constraint from the CHOOZ experiment makes it
more difficult to observe CP violation in neutrino oscillation
experiments. But a closer examination reveals that the situation is
not so simple and in fact it depends upon the scenarios of neutrino
mass hierarchies.  In the ADN or the ELN type mass hierarchies, the
estimation of the magnitude of CP violation in long-baseline neutrino
oscillation experiments done by Arafune, Koike and Sato \cite {AKS}
remains unchanged because the (pin-pointed) mixing parameters they
used is consistent with the bound from the CHOOZ data. In the HMN
case, however, we cannot allow all the $s^2_{12}$ regions analyzed in
Refs. \cite {MN}, but should restrict to the region $s^2_{12} \leq
0.05$, assuming that $\Delta m^2_{\rm atm} \geq 2 \times 10^{-3}$
eV$^2$. The magnitude of the CP violating effect thereby decreases
depending upon the value of other mixing angles.

\section*{ACKNOWLEDGEMENT}
This research was supported in part by a Grant-in-Aid for Scientific
Research of the Ministry of Education, Science and Culture,
\#09640370, and 
by Grant-in-Aid for Scientific Research \#09045036 
under International Scientific Research Program, Inter-University 
Cooperative Research.

\newpage

\vspace{1.5cm}
\centerline{\large FIGURE CAPTIONS}
\vspace{0.5cm}

\begin{description}

\item[Fig.1]
\begin{minipage}[t]{13cm}
\baselineskip=20pt
A schematic illustration of the mass pattern with hierarchy 
in differences of squared masses discussed in this paper. 
We deal with the two mass patterns (A) and (B) simultaneously while 
they can be distinguished by the matter effect in the earth. 
\end{minipage}

\item[Fig.2]
\begin{minipage}[t]{13cm}
\baselineskip=20pt
The allowed region ($\Delta m^2$, $\sin^2 2\theta$) in a framework of
two flavor neutrino oscillations obtained by combining the CHOOZ
\cite{CHOOZ}, the Bugey \cite{Bugey}, and the Krasnoyarsk
\cite{Krasnoyarsk} experiments. The abscissa can be regarded as
$\sin^2 2\theta_{12}$ in the HMN scenario in its parameter region (a),
as discussed in the text.
\end{minipage}

\end{description}

\pagestyle{empty}
\newpage
\epsfig{file=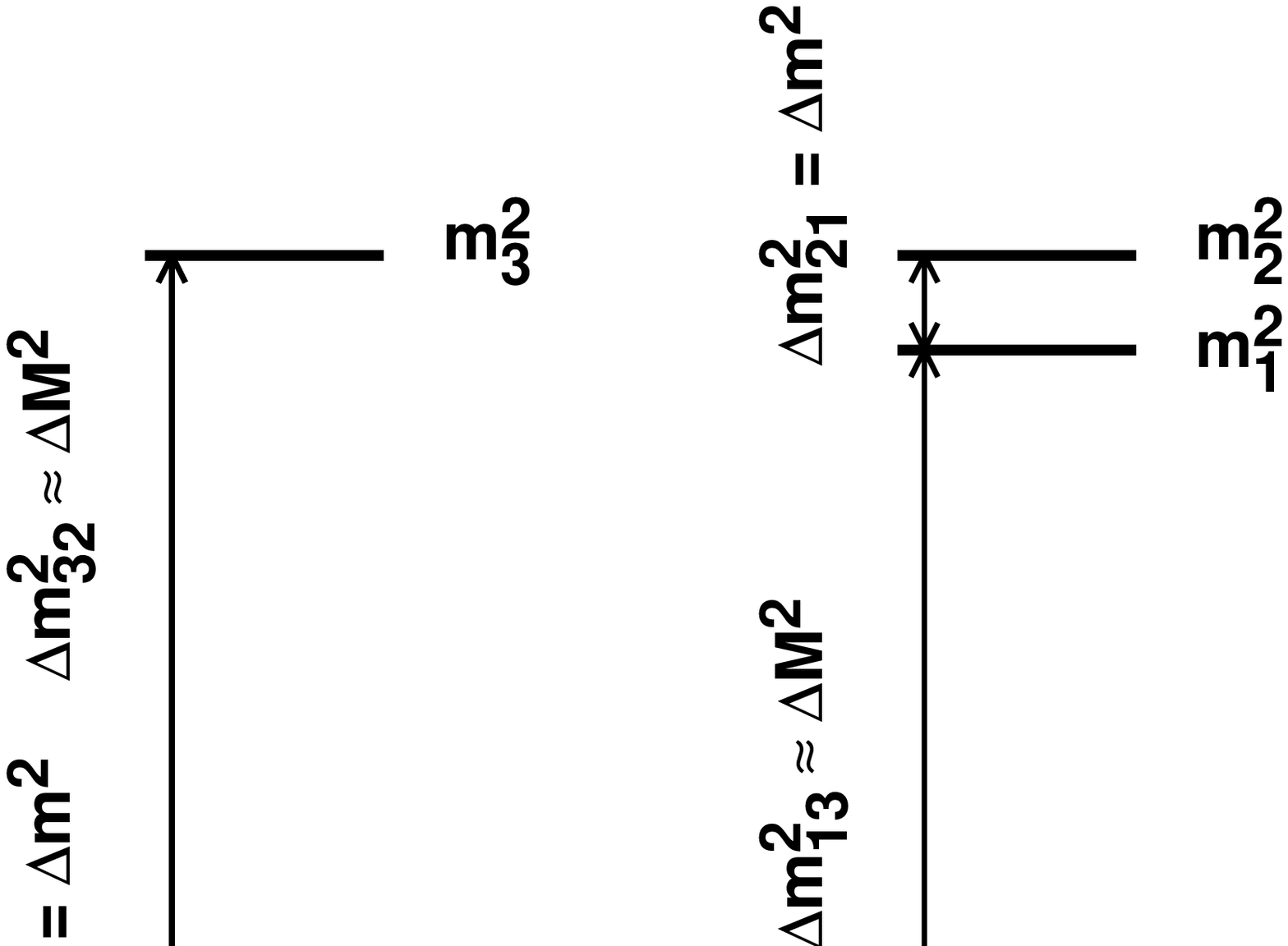,width=15cm}
\newpage
\epsfig{file=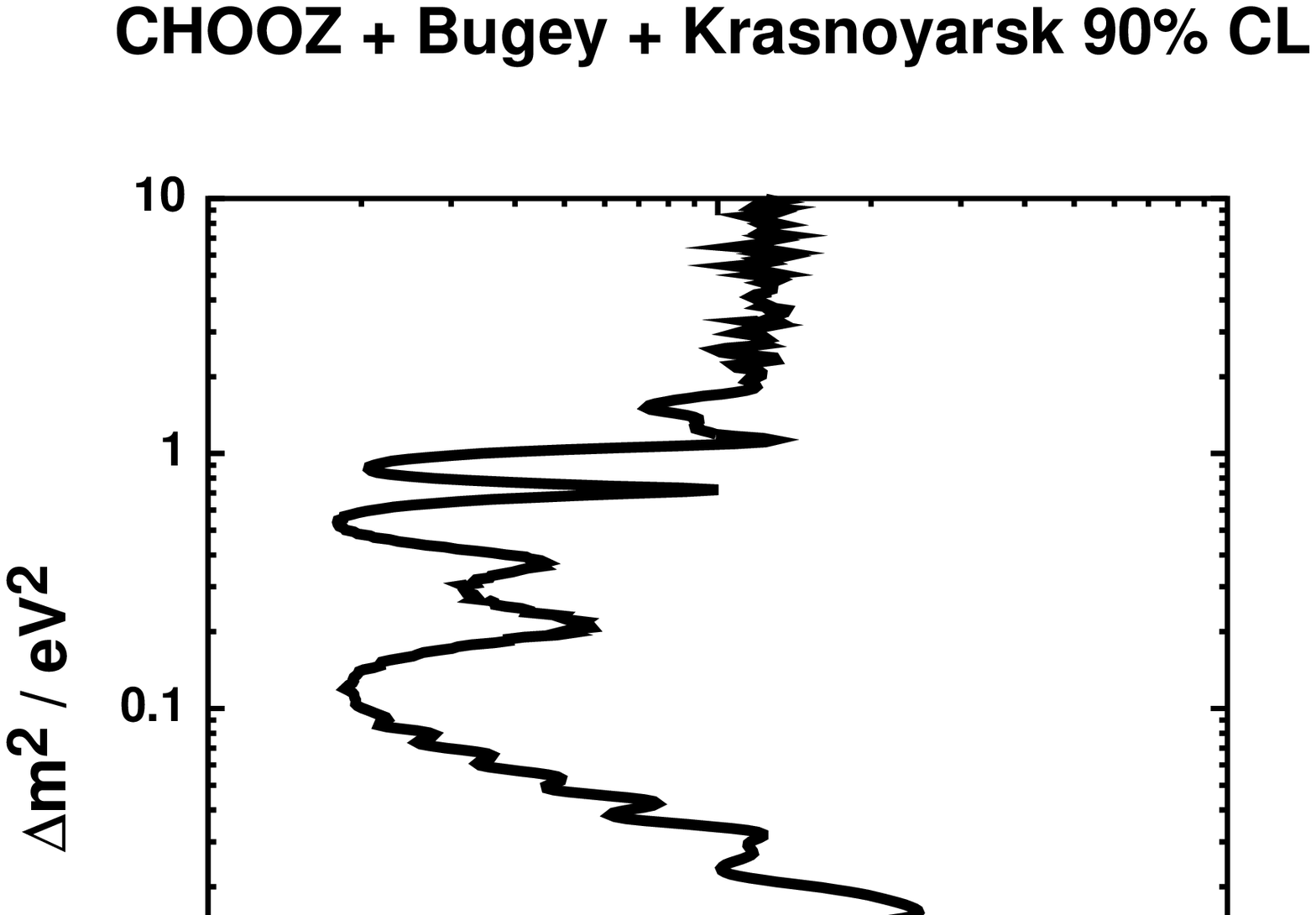,width=15cm}

\end{document}